\documentclass[aps,floats]{revtex4}
\usepackage{amsmath,amssymb}
\usepackage{graphicx,epsfig}
\usepackage[greek,english]{babel}

\begin{document}
\bibliographystyle {plain}

\def\oppropto{\mathop{\propto}} 
\def\opsimeq{\mathop{\simeq}}
\def\opoverderline{\mathop{\overline}}
\def\operarrow{\mathop{\longrightarrow}}
\def\opsim{\mathop{\sim}} 
\def\opmin{\mathop{\min}} 
\def\opmax{\mathop{\max}} 

\def\fig#1#2{\includegraphics[height=#1]{#2}}
\def\figx#1#2{\includegraphics[width=#1]{#2}}


\title{Boundary-driven Lindblad dynamics of random quantum spin chains : \\
strong disorder approach for the relaxation, the steady state and the current } 


\author{ C\'ecile Monthus }
 \affiliation{Institut de Physique Th\'{e}orique, 
Universit\'e Paris Saclay, CNRS, CEA,
91191 Gif-sur-Yvette, France}

\begin{abstract}
The Lindblad dynamics of the XX quantum chain with large random fields $h_j$ (the couplings $J_j$ can be either uniform or random) is considered for boundary-magnetization-drivings acting on the two end-spins. Since each boundary-reservoir tends to impose its own magnetization, we first study the relaxation spectrum in the presence of a single reservoir as a function of the system size via some boundary-strong-disorder renormalization approach. The non-equilibrium-steady-state in the presence of two reservoirs can be then analyzed from the effective renormalized Linbladians associated to the two reservoirs. The magnetization is found to follow a step profile, as found previously in other localized chains. The strong disorder approach allows to compute explicitly the location of the step of the magnetization profile and the corresponding magnetization-current for each disordered sample in terms of the random fields and couplings.

\end{abstract}

\maketitle

\section{ Introduction }

In the field of random quantum spin chains, the interplay of disorder and dissipation
has attracted a lot of attention recently. As a first example, 
the coupling to a dissipative bath of harmonic oscillators
with some spectral function as in the spin-boson model \cite{spinboson}
has been analyzed via Strong Disorder Renormalization \cite{greg_rieger1,greg_rieger2,hoyos1,hoyos2,hoyos3,hoyos4,hoyos5,hoyos6}.
As a second example, the Lindblad dynamics with boundary-driving and/or dephasing 
has been studied for Many-Body-Localization models in various regimes
\cite{garrahan_mbl,prosen_mbl,altman_mbl,znidaric_mblsubdiff,znidaric_mbldeph,huveneers_mblstep}.

Among the various descriptions of open quantum systems \cite{open}, 
one of the most effective is indeed
the Lindblad equation for the density matrix $\rho(t)$
\begin{eqnarray}
\frac{\partial \rho(t) }{\partial t} = {\cal L}[\rho(t) ]= {\cal U}[\rho(t) ]  +  {\cal D}[\rho(t)]
\label{dynlindblad}
\end{eqnarray}
where the Lindblad operator $ {\cal L}$ contains the unitary evolution as if the system of Hamiltonian $H$ were isolated
\begin{eqnarray}
 {\cal U}[\rho(t) ] \equiv -i [H,\rho(t) ] 
\label{unitary}
\end{eqnarray}
and the dissipative contribution defined in terms of some set of operators $L_{\alpha}$ that describe the interaction with the reservoirs
(see example in section II)
\begin{eqnarray}
 {\cal D}[\rho(t)] = \sum_{\alpha} \gamma_\alpha \left(L_\alpha \rho(t) L_\alpha^{\dagger} - \frac{1}{2} L_\alpha^{\dagger}L_\alpha \rho(t)- \frac{1}{2} \rho(t) L_\alpha^{\dagger}L_\alpha \right)
\label{dissi}
\end{eqnarray}
so that the trace of the density matrix is conserved by the dynamics
\begin{eqnarray}
\frac{\partial  }{\partial t} {\rm Tr} (\rho(t) ) =0
\label{conservtrace}
\end{eqnarray}
The first advantage of this formulation of the dynamics as a Quantum Markovian Master Equation 
is that the relaxation properties can be studied from the spectrum of the Lindblad operator
\cite{cai_barthel,kollath,znidaric_relaxation}
with possible metastability phenomena \cite{garrahan_metastability}.
This spectral analysis also allows to make some link with the Random Matrix Theory of eigenvalues statistics \cite{link_RMTstatistics}.
The second advantage is that this framework is very convenient
to study the non-equilibrium transport properties  
\cite{znidaric_deph,znidaric_trans,znidaric_heisen,hubbard,horvat,clark_heat,clark_step,prosen_step,dragi_trans}
with many exact solutions 
\cite {prosen_third,znidaric_solvable,dragi_ex,dragi_twist,prosen_mpsolu,dragi_mps}.
In addition, 
many important ideas that have been developed in the context of classical non-equilibrium systems
(see the review \cite{derrida} and references therein)
have been adapted to the Lindblad description of non-equilibrium dissipative quantum systems,
in particular the large deviation formalism to access the full-counting statistics 
\cite{garrahan_s,ates_s,hickey_s,genway_s,znidaric_slargedev,znidaric_sanomalous,prosen_s,pigeon},
the additivity principle \cite{znidaric_sadditivity}
and the fluctuation relations \cite{chetrite}.

In the present paper, we consider the XX chain of $N$ spins with random fields $h_j$ and
 couplings $J_j$ (that can be either uniform or random)
\begin{eqnarray}
H&& = \sum_{j=1}^N \left[ h_j \sigma_j^z + J_j ( \sigma_j^x \sigma_{j+1}^x+\sigma_j^y \sigma_{j+1}^y ) \right]
 = \sum_{j=1}^N \left[ h_j \sigma_j^z
 + 2 J_j ( \sigma_j^+ \sigma_{j+1}^-+\sigma_j^- \sigma_{j+1}^+ ) \right]
\label{XXh}
\end{eqnarray}
and analyze the Lindblad dynamics in the presence of two boundary-magnetization-drivings
acting on the two end-spins. We focus
on the strong disorder regime where the scale of the random fields $(h_j)$ is much bigger than the couplings $(J_j)$.
The paper is organized as follows.
In section \ref{sec_lindblad}, we introduce the notations for the boundary-magnetization-drivings
and we recall the spectral analysis of the Lindbladian in the ladder formulation,
as well as the notion of tilted-Lindbladian to access the full-counting statistics
of the exchanges with one reservoir.
In section \ref{sec_twospins}, we show how this formalism works in practice on the simplest case of $N=2$ spins. We then turn to the case of a chain of arbitrary length $N$ :
in section \ref{sec_onereservoir}, we analyze the relaxation properties of a long chain 
in contact with a single reservoir, while in section \ref{sec_tworeservoirs},
we analyze the non-equilibrium-steady-state for the chain coupled to two reservoirs :
the magnetization profile and the magnetization current are computed in the strong disorder regime.
Our conclusions are summarized in section \ref{sec_conclusion}.

\section{ Lindblad dynamics with boundary-magnetization-driving }

\label{sec_lindblad}

\subsection{ Boundary-magnetization-driving on the end-spins $\sigma_1$ and $\sigma_N$ }

The standard boundary-magnetization-driving on the first spin $\sigma_1$
is based on the dissipative operator of Eq. \ref{dissi} with the two operators $\alpha=1,2$
\begin{eqnarray}
L_1 && = \sigma_1^+ 
\nonumber \\
L_2 && = \sigma_1^- 
\label{4Lboundary}
\end{eqnarray}
and the corresponding amplitudes
\begin{eqnarray}
\gamma_1 && =\Gamma \frac{1+\mu}{2} 
\nonumber \\
\gamma_2 && = \Gamma \frac{1-\mu}{2} 
\label{4gboundary}
\end{eqnarray}
leading to
\begin{eqnarray}
 {\cal D}^{spin1}[\rho] && = \Gamma \frac{1+\mu}{2}  \left( \sigma_1^+ \rho \sigma_1^-  - \frac{1}{2} \sigma_1^- \sigma_1^+ \rho- \frac{1}{2} \rho  \sigma_1^- \sigma_1^+\right)
\nonumber \\
&& 
+\Gamma\frac{1-\mu}{2}  \left( \sigma_1^- \rho \sigma_1^+  - \frac{1}{2} \sigma_1^+ \sigma_1^- \rho- \frac{1}{2} \rho  \sigma_1^+ \sigma_1^-\right)
\label{dissiboundaryleft}
\end{eqnarray}
Using the identities
\begin{eqnarray}
\sigma_1^- \sigma_1^+ && = \frac{1-\sigma_1^z}{2}
\nonumber \\
\sigma_1^+ \sigma_1^- && = \frac{1+\sigma_1^z}{2}
\label{pauliid}
\end{eqnarray}
Eq \ref{dissiboundaryleft} becomes
\begin{eqnarray}
 {\cal D}^{spin1}[\rho] && = 
\Gamma \left( \frac{1+\mu}{2}  \sigma_1^+ \rho \sigma_1^- + \frac{1-\mu}{2}  \sigma_1^- \rho \sigma_1^+  \right)
-  \frac{ \Gamma}{2} \rho
 + \frac{ \Gamma \mu}{4} (\sigma_1^z \rho + \rho \sigma_1^z )
\label{dissiboundaryleftres}
\end{eqnarray}
The physical meaning of this dissipative operator is that it tends to
impose the magnetization $(+\mu)$ on the spin 1 with a characteristic relaxation rate of order $\Gamma$.

A simple way to generate a non-equilibrium steady-state
is to consider 
a similar boundary-magnetization-driving on the last spin $\sigma_N$
that tend to impose another magnetization $\mu' \ne \mu $ with some rate $\Gamma'$,
so that the corresponding dissipative operator reads
\begin{eqnarray}
 {\cal D}^{spinN}[\rho] && = 
\Gamma' \left( \frac{1+\mu'}{2}  \sigma_N^+ \rho \sigma_N^- + \frac{1-\mu'}{2}  \sigma_N^- \rho \sigma_N^+  \right)
-  \frac{ \Gamma'}{2} \rho
 + \frac{ \Gamma' \mu'}{4} (\sigma_N^z \rho + \rho \sigma_N^z )
\label{dissiboundaryright}
\end{eqnarray}

\subsection{ Ladder Formulation of the Lindbladian}

Since the Lindblad operator acts on the density matrix $\rho(t)$ of the chain of $N$ spins
that can be expanded in the $\sigma^z$ basis 
\begin{eqnarray}
\rho(t) = \sum_{S_1=\pm 1} ...
 \sum_{S_N=\pm 1}  \sum_{T_1=\pm 1} ... \sum_{T_N=\pm 1} 
  \rho_{S_1,..,S_N;T_1,...,T_N} (t) \vert S_1,...,S_N > < T_1,...,T_N \vert
\label{rhoexpcoefs}
\end{eqnarray}
in terms of the $4^N$ coefficients
\begin{eqnarray}
 \rho_{S_1,..,S_N;T_1,...,T_N } (t) = <  S_1,...,S_N \vert \rho(t) \vert T_1,...,T_N >
\label{rhocoefs}
\end{eqnarray}
it can be technically convenient to 'vectorize' the density {\it matrix } 
of the {\it spin chain } \cite{znidaric_relaxation,znidaric_sadditivity,jakob,savona,cirac},
i.e. to consider that these $4^N$ coefficients
are the components of a { \it ket } describing the state of a { \it spin ladder }
\begin{eqnarray}
\vert \rho (t) >^{Ladder} = \sum_{S_1=\pm 1} ... \sum_{S_N=\pm 1}  \sum_{T_1=\pm 1} ... \sum_{T_N=\pm 1} 
  \rho_{S_1,..,S_N;T_1,...,T_N} (t) \vert S_1,...,S_N > \otimes \vert T_1,...,T_N >
\label{rholadder}
\end{eqnarray}

To translate the Lindblad operator of Eq. \ref{dynlindblad}
 in this ladder formulation, one needs to  
consider the product $(A \rho(t) B)$ where $A$ and $B$ are two arbitrary matrices 
\begin{eqnarray}
&& A \rho(t) B  = \sum_{S_1=\pm 1} ... \sum_{S_N=\pm 1}  \sum_{T_1=\pm 1} ... \sum_{T_N=\pm 1} 
  \rho_{S_1,..,S_N;T_1,...,T_N} (t) A  \vert S_1,...,S_N > < T_1,...,T_N \vert B
\nonumber \\
&& =
\sum_{S_1'=\pm 1} ... \sum_{S_N'=\pm 1}  \sum_{T_1'=\pm 1} ... \sum_{T_N'=\pm 1} 
\vert S_1',...,S_N' >< T_1',...,T_N'  \vert 
\nonumber \\
&& \sum_{S_1=\pm 1} ... \sum_{S_N=\pm 1}  \sum_{T_1=\pm 1} ... \sum_{T_N=\pm 1} 
  < S_1',...,S_N'  \vert A  \vert S_1,...,S_N > \rho_{S_1,..,S_N;T_1,...,T_N} (t) 
 < T_1,...,T_N \vert B \vert T_1',...,T_N' > 
\label{2matrix}
\end{eqnarray}
and to write the corresponding ket
\begin{eqnarray}
&& \vert A \rho(t) B  >^{Ladder}
 =
\sum_{S_1'=\pm 1} ... \sum_{S_N'=\pm 1}  \sum_{T_1'=\pm 1} ... \sum_{T_N'=\pm 1} 
\vert S_1',...,S_N' > \otimes \vert T_1',...,T_N' > 
\nonumber \\
&& \sum_{S_1=\pm 1} ... \sum_{S_N=\pm 1}  \sum_{T_1=\pm 1} ... \sum_{T_N=\pm 1} 
  < S_1',...,S_N'  \vert A  \vert S_1,...,S_N > \rho_{S_1,..,S_N;T_1,...,T_N} (t) 
 < T_1',...,T_N' \vert B^T \vert T_1,...,T_N > 
\nonumber \\
&& = A \otimes B^T  \vert \rho (t) >^{Ladder} 
\label{2matrixket}
\end{eqnarray}
where $B^T$ denotes the transpose of the matrix $B$.
As a consequence, the Lindblad operator governing the evolution of the ket $\vert \rho (t) >^{Ladder} $
\begin{eqnarray}
\frac{\partial \vert \rho (t) >^{Ladder} }{\partial t} = {\cal L}^{Ladder}
\vert \rho (t) >^{Ladder}
\label{dynlindbladladder}
\end{eqnarray}
can be translated from Eqs \ref{unitary} and \ref{dissi} and reads
\begin{eqnarray}
 {\cal L}^{Ladder} &&  = -i  (H \otimes \mathbb{I}-\mathbb{I}\otimes H^T)+ 
 \sum_\alpha \gamma_\alpha 
\left( L_\alpha \otimes (L_\alpha^{\dagger})^T 
-  \frac{1}{2} L_\alpha^{\dagger} L_\alpha \otimes \mathbb{I}
-  \frac{1}{2}\mathbb{I}\otimes (L_\alpha^{\dagger} L_\alpha)^T  \right)
\nonumber \\
&&  = -i  (H \otimes \mathbb{I}-\mathbb{I}\otimes H)+ 
 \sum_{\alpha} \gamma_\alpha 
\left( L_\alpha \otimes L_\alpha^{*}
-  \frac{1}{2} L_\alpha^{\dagger} L_\alpha \otimes \mathbb{I}
-  \frac{1}{2}\mathbb{I}\otimes L_\alpha^{T} L_\alpha^*  \right)
\label{lindbladladder}
\end{eqnarray}

For the chain of Eq. \ref{XXh}, the unitary part reads
in terms of the Pauli matrices of the spin ladder
\begin{eqnarray}
 {\cal U}^{Ladder}  
&& = -i  \sum_{j=1}^N \left[ h_j \sigma_j^z 
 + 2 J_j ( \sigma_j^+ \sigma_{j+1}^-+\sigma_j^- \sigma_{j+1}^+ ) \right] 
\nonumber \\
&&+ i  \sum_{j=1}^N \left[ h_j \tau_j^z 
 + 2 J_j ( \tau_j^+ \tau_{j+1}^-+\tau_j^- \tau_{j+1}^+ ) \right] 
\label{uladder}
\end{eqnarray}
while the dissipative operators of Eqs \ref{dissiboundaryleft} and \ref{dissiboundaryright}
become
\begin{eqnarray}
 {\cal D}^{Ladder}_{Spin1} && = 
\Gamma \left( \frac{1+\mu}{2}  \sigma_1^+ \tau_1^+ + \frac{1-\mu}{2}  \sigma_1^- \tau_1^-  \right)
-  \frac{ \Gamma}{2}
 + \frac{ \Gamma \mu}{4} (\sigma_1^z + \tau_1^z )
\label{dissiboundaryleftladder}
\end{eqnarray}
and
\begin{eqnarray}
 {\cal D}^{Ladder}_{SpinN} && = 
\Gamma' \left( \frac{1+\mu'}{2}  \sigma_N^+ \tau_N^+ + \frac{1-\mu'}{2}  \sigma_N^- \tau_N^-  \right)
-  \frac{ \Gamma' }{2}
 + \frac{ \Gamma' \mu'}{4} (\sigma_N^z + \tau_N^z )
\label{dissiboundaryrightladder}
\end{eqnarray}

\subsection { Spectral Decomposition of the Ladder Lindbladian }

The ladder formulation of the Lindbladian described above
is especially useful to use the very convenient bra-ket notations
to denote the Right and Left eigenvectors
associated to the $4^N$ eigenvalues $\lambda_n$
\begin{eqnarray}
 {\cal L}^{Ladder}  \vert \psi^R_{\lambda_n} > && = \lambda_n  \vert \psi^R_{\lambda_n} >
\nonumber \\
<\psi^L_{\lambda_n} \vert {\cal L}^{Ladder}   && = \lambda_n <\psi^L_{\lambda_n} \vert
\label{spectraleigen}
\end{eqnarray}
with the orthonormalization
\begin{eqnarray}
  <\psi^L_{\lambda_n} \vert  \psi^R_{\lambda_m} > = \delta_{nm}
\label{orthonorm}
\end{eqnarray}
and the identity decomposition
\begin{eqnarray}
1 = \sum_{n=0}^{4^N-1}   \vert \psi^R_{\lambda_n} >  <\psi^L_{\lambda_n} \vert 
\label{identity}
\end{eqnarray}
The spectral decomposition of the Lindbladian
\begin{eqnarray}
 {\cal L}^{Ladder} =  \sum_{n=0}^{4^N-1}  \lambda_n   \vert \psi^R_{\lambda_n} >  <\psi^L_{\lambda_n} \vert 
\label{spectral}
\end{eqnarray}
then allows to write the solution for the dynamics in terms of the initial condition at $t=0$ as
\begin{eqnarray}
\vert \rho^{Ladder}(t) >=  \sum_{n=0}^{4^N-1}  e^{\lambda_n t}   \vert \psi^R_{\lambda_n} > 
 <\psi^L_{\lambda_n} \vert \rho^{Ladder}(t=0) >
\label{spectralrlax}
\end{eqnarray}

The trace of the density matrix $\rho(t)$ corresponds  in the Ladder Formulation to
\begin{eqnarray}
{\rm Tr }(\rho(t) ) =\sum_{S_1=\pm 1} ... \sum_{S_N=\pm 1} 
  \rho_{S_1,..,S_N;S_1,...,S_N} (t)
=  \sum_{S_1=\pm 1} ... \sum_{S_N=\pm 1} <  S_1,..,S_N \vert \otimes <S_1,...,S_N
\vert \rho (t) >^{Ladder}
\label{tracerholader}
\end{eqnarray}
Its conservation by the dynamics (Eq \ref{conservtrace}) means that the eigenvalue
\begin{eqnarray}
\lambda_0=0
\label{zeroeigen}
\end{eqnarray}
is associated to the Left eigenvector
\begin{eqnarray}
  <\psi^L_{\lambda_0=0} \vert  = \sum_{S_1=\pm 1} ... \sum_{S_N=\pm 1}
 <  S_1,..,S_N \vert \otimes <S_1,...,S_N
\vert
\label{lefteigenzero}
\end{eqnarray}
while the corresponding Right Eigenvector corresponds to the steady state towards which any initial condition will converges
\begin{eqnarray}
\vert \rho^{Ladder}(t\to +\infty ) >=    \vert \psi^R_{\lambda_0=0} > 
\label{righteigenzero}
\end{eqnarray}
The other $(4^N-1)$ eigenvalues $\lambda_{n \ne 0}$ with negative real parts
describe the relaxation towards this steady state.

\subsection{ Tilted-Lindbladian $ {\cal L}(s)  $ to measure the exchanges with the boundary-reservoir on spin 1} 

As mentioned in the Introduction, the method of 'tilting' the master equation
to access the full-counting statistics developed for classical non-equilibrium models 
(see the review \cite{derrida} and references therein)
has been adapted to the Lindblad framework
\cite{garrahan_s,ates_s,hickey_s,genway_s,znidaric_slargedev,znidaric_sanomalous,prosen_s,pigeon,znidaric_sadditivity}
as follows.
To keep the information on the global number $N_t$ of 'magnetization particles' that have been exchanged with the reservoir 
acting on the spin 1 since the initial condition at $t=0$,
it is convenient to decompose the Lindbladian into three terms
\begin{eqnarray}
 {\cal L}^{Ladder}=  {\cal L}^{Ladder}_0+ {\cal L}^{Ladder}_{+}+ {\cal L}^{Ladder}_{-}
\label{lindblad0pm}
\end{eqnarray}
where 
\begin{eqnarray}
 {\cal L}^{Ladder}_{+}  && = \Gamma  \frac{1+\mu}{2}  \sigma_1^+ \tau_1^+ 
\nonumber \\
 {\cal L}^{Ladder}_{-} && =\Gamma  \frac{1-\mu}{2}  \sigma_1^- \tau_1^-  
\label{lindbladpm}
\end{eqnarray}
describe respectively the processes corresponding to an increase ($N_t \to N_t+1$) and a decrease ($N_t \to N_t-1$)
 by an elementary 'magnetization particle', while $ {\cal L}^{Ladder}_0 $ 
contains all the other terms of the Lindbladian that do not correspond to an exchange
with the reservoir acting on spin 1 ($N_t \to N_t$).
As a consequence, 
the eigenvalue $\lambda_0(s)$ with the largest real-part
of the tilted-Lindbladian by the parameter $s$
\begin{eqnarray}
 {\cal L}^{Ladder}(s) =  {\cal L}^{Ladder}_0+ e^{s}{\cal L}^{Ladder}_{+}+ e^{-s} {\cal L}_{-}^{Ladder}
\label{lindblads}
\end{eqnarray}
allows to obtain the statistics of the number $N_t$ in the large-time regime via
\begin{eqnarray}
\lambda_0(s) = \lim \limits_{t \to +\infty} \frac{\ln < e^{s N_t} > }{ t} 
\label{eigens}
\end{eqnarray}
In particular, the expansion up to second order in $s$ 
\begin{eqnarray}
\lambda_0(s) = s I_{av} + \frac{s^2}{2} F +O(s^3)
\label{eigensexpansion}
\end{eqnarray}
gives the averaged current entering from the reservoir acting on the spin 1
\begin{eqnarray}
I_{av}  = \lim \limits_{t \to +\infty} \frac{ <N_t> }{ t} 
\label{siav}
\end{eqnarray}
and the fluctuation
\begin{eqnarray}
F  = \lim \limits_{t \to +\infty} \frac{ (<N_t^2>-<N_t>^2 ) }{ t} 
\label{sfluct}
\end{eqnarray}

More generally, the whole large-deviation properties of the probability distribution $P_t(I)$ of the current $I=\frac{N_t}{t}$
\begin{eqnarray}
P_t(I) \opsimeq_{t \to +\infty} e^{-t  \Phi(I) }
\label{larfedev}
\end{eqnarray}
can be obtained as the Legendre transform of the tilted eigenvalue of Eq. \ref{eigens}
\begin{eqnarray}
  \Phi(I) = \opmax_s ( s I - \lambda_0(s) )
\label{legendre}
\end{eqnarray}

\subsection{Notation  }

In the remaining of this paper, the ladder formulation of the Lindblad operator described above will be always used,
so that the explicit mention 'Ladder' will be dropped from now on in order to simplify the notations.

\section{  Strong-Disorder Approach for $N=2$ spins }

\label{sec_twospins}

To see how the formalism recalled in the previous section
 works in practice, it is 
useful to focus first on the simplest example of $N=2$ spins.
In addition, to motivate the Strong-Disorder approach
for long chains $N \gg 1$ that will be described in the following sections,
we will consider that the only term of the Linbladian
that couples the two spins
\begin{eqnarray}
 {\cal L}^{per (1,2)}
=
&& i   2 J_1 ( \tau_1^+ \tau_{2}^-+\tau_1^- \tau_{2}^+ - \sigma_1^+ \sigma_{2}^--\sigma_1^- \sigma_{2}^+)
\label{pertwo}
\end{eqnarray}
is a perturbation with respect to all the other terms that do not couple the two spins
\begin{eqnarray}
 {\cal L}^{unper} && =  {\cal L}^{spin1}(s)  +  {\cal L}^{spin2}
\label{unpertwo}
\end{eqnarray}

\subsection { Spectral decomposition of ${\cal L}^{spin1}(s) $   }

The tilted Lindbladian of Eq. \ref{lindblad0pm} for the spin 1
\begin{eqnarray}
 {\cal L}^{spin1}(s) =  i h_1 (\tau_1^z-\sigma_1^z)
-  \frac{ \Gamma}{2}
 + \frac{ \Gamma \mu}{4} (\sigma_1^z + \tau_1^z )
\nonumber + e^{s}  \Gamma  \frac{1+\mu}{2}  \sigma_1^+ \tau_1^+ 
+ e^{-s}  \Gamma  \frac{1-\mu}{2}  \sigma_1^- \tau_1^-  
\label{lindbladspin1}
\end{eqnarray}
has the following four eigenvalues that do not depend on the tilting parameter $s$
in contrast to some corresponding eigenvectors written in the basis $(\sigma_1^z,\tau^z_1)$:

(0) The eigenvalue $\lambda_{n=0}^{spin1}(s) =0$ is associated to 
\begin{eqnarray}
  <  \lambda_{n=0}^{spin1 (L)} (s)\vert  && =  e^{-s} < ++ \vert + < -- \vert
\nonumber \\
    \vert \lambda_{n=0}^{spin1 (R)} (s) > && = e^{s} \frac{1+\mu}{2} \vert ++> +\frac{1-\mu}{2} \vert -- >  
\label{lindbladspin1n0}
\end{eqnarray}

(1) The eigenvalue $\lambda_{n=1}^{spin1}(s)=-\Gamma$ is associated to 
\begin{eqnarray}
  < \lambda_{n=1}^{spin1 (L)}(s) \vert  && = e^{-s} \frac{1-\mu}{2} < ++ \vert  - \frac{1+\mu}{2} <--\vert 
\nonumber \\
    \vert  \lambda_{n=1}^{spin1 (R)}(s) > && = e^{s} \vert ++>  - \vert -->
\label{lindbladspin1n1}
\end{eqnarray}

(2) The eigenvalue $\lambda_{n=2}^{spin1}(s)=-\frac{\Gamma}{2}+i 2h_1$ is associated to 
\begin{eqnarray}
  <\lambda_{n=2}^{spin1 (L)}(s) \vert  && = < -+ \vert  
\nonumber \\
    \vert \lambda_{n=2}^{spin1 (R)}(s) > && = \vert -+>  
\label{lindbladspin1n2}
\end{eqnarray}

(4) The eigenvalue $\lambda_{n=3}^{spin1}(s)=-\frac{\Gamma}{2}-i 2h_1$ is associated to 
\begin{eqnarray}
  < \lambda_{n=3}^{spin1 (L)}(s)\vert  && = < +- \vert  
\nonumber \\
    \vert \lambda_{n=3}^{spin1 (R)}(s) > && = \vert +->  
\label{lindbladspin1n3}
\end{eqnarray}

\subsection { Spectral decomposition of ${\cal L}^{spin2} $   }

The non-tilted Lindbladian for the spin $N=2$
\begin{eqnarray}
 {\cal L}^{spin2} =  i h_2 (\tau_2^z-\sigma_2^z)
-  \frac{ \Gamma'}{2}
 + \frac{ \Gamma' \mu'}{4} (\sigma_2^z + \tau_2^z )
\nonumber +  \Gamma'  \frac{1+\mu'}{2}  \sigma_2^+ \tau_2^+ 
+   \Gamma'  \frac{1-\mu'}{2}  \sigma_2^- \tau_2^-  
\label{lindbladspin2}
\end{eqnarray}
has the following four eigenvalues and eigenvectors in the basis $(\sigma_2^z,\tau^z_2)$ :

(0) The eigenvalue $\lambda_{m=0}^{spin2} =0$ is associated to 
\begin{eqnarray}
  <  \lambda_{m=0 }^{spin2 (L)} \vert  && =   < ++ \vert + < -- \vert
\nonumber \\
    \vert \lambda_{m=0 (R)}^{spin2}  > && = \frac{1+\mu'}{2} \vert ++> +\frac{1-\mu'}{2} \vert -- >  
\label{lindbladspin2n0}
\end{eqnarray}

(1) The eigenvalue $\lambda_{m=1}^{spin2}=-\Gamma'$ is associated to 
\begin{eqnarray}
  < \lambda_{m=1}^{spin2 (L)} \vert  && =  \frac{1-\mu'}{2} < ++ \vert  - \frac{1+\mu'}{2} <--\vert 
\nonumber \\
    \vert  \lambda_{m=1}^{spin2 (R)} > && =  \vert ++>  - \vert -->
\label{lindbladspin2n1}
\end{eqnarray}

(2) The eigenvalue $\lambda_{m=2}^{spin2}=-\frac{\Gamma'}{2}+i 2h_2$ is associated to 
\begin{eqnarray}
  <\lambda_{m=2}^{spin2 (L)} \vert  && = < -+ \vert  
\nonumber \\
    \vert \lambda_{m=2}^{spin2 (R)} > && = \vert -+>  
\label{lindbladspin2n2}
\end{eqnarray}

(4) The eigenvalue $\lambda_{n=3}^{spin2}=-\frac{\Gamma'}{2}-i 2h_2$ is associated to 
\begin{eqnarray}
  < \lambda_{m=3}^{spin2 (L)}\vert  && = < +- \vert  
\nonumber \\
    \vert \lambda_{m=3}^{spin2 (R)} > && = \vert +->  
\label{lindbladspin2n3}
\end{eqnarray}

\subsection { Second-Order perturbation theory in the coupling $ {\cal L}^{per(1,2)} $    }

The unperturbed Lindbladian of Eq. \ref{unpertwo} is the sum of the two independent Lindbladians discussed above,
so its 16 eigenvalues are simply given by the sum of eigenvalues for $n=0,1,2,3$ and $m=0,1,2,3$
\begin{eqnarray}
 \lambda^{unper}_{n,m} && =\lambda_{n}^{spin1}+ \lambda_{m}^{spin2}
\label{unpertwonm}
\end{eqnarray}
while the left and right eigenvectors are given by the corresponding tensor-products 
\begin{eqnarray}
  < \lambda^{unper (L)}_{n,m}  \vert  && =   < \lambda_{n}^{spin1 (L)}\vert \otimes   < \lambda_{m}^{spin2 (L)}\vert 
\nonumber \\
    \vert  \lambda^{unper (R)}_{n,m} > && =   \vert \lambda_{n}^{spin1 (R)} > \otimes   \vert \lambda_{m}^{spin2 (R)} >
\label{unpertwovec}
\end{eqnarray}

Here we are interested into the eigenvalue $\lambda_{0}(s)$ with the largest real part 
of the tilted Lindbladian (Eq. \ref{lindblads}).
The corresponding unperturbed eigenvalue vanishes
\begin{eqnarray}
 \lambda^{unper}_{n=0,m=0}(s) && = 0
\label{unpertwozero}
\end{eqnarray}
but it will become non-zero and depend on the parameter $s$
when the coupling between the two spins is taken into account
by the second-order perturbation theory
\begin{eqnarray}
\lambda_{0}(s) = 
\sum_{(n,m) \ne (0,0 )} 
\frac{ < \lambda^{unper (L)}_{0,0}   \vert
{\cal L}^{per (1,2)} 
 \vert \lambda^{unper (R)}_{n,m}   >
< \lambda^{unper (L)}_{n,m}   \vert
{\cal L}^{per (1,2)} 
\vert \lambda^{unper (R)}_{0,0} > }
{  \lambda^{unper}_{0,0} -  \lambda^{unper}_{n,m} } 
\label{nondegeeigenpertwo}
\end{eqnarray}

The application of the perturbation ${\cal L}^{per (1,2)} $
to the left unperturbed eigenvector
\begin{eqnarray}
  < \lambda^{unper (L)}_{0,0}  \vert {\cal L}^{per (1,2)}  && = 
i 2 J_1 ( e^{-s} -1) ( < \lambda^{unper (L)}_{3,2}  \vert - < \lambda^{unper (L)}_{2,3}  \vert ) 
\label{appliper12l}
\end{eqnarray}
and to the right unperturbed eigenvector
\begin{eqnarray}
 {\cal L}^{per (1,2)}    \vert  \lambda^{unper (R)}_{n,m} > && =  
i 2 J_1 \frac{ e^s (1+\mu)(1-\mu') - (1-\mu)(1+\mu') }{4} ( \vert \lambda^{unper (L)}_{3,2}  >- \vert \lambda^{unper (L)}_{2,3}  > ) 
\label{appliper12r}
\end{eqnarray}
shows that the formula of Eq. \ref{nondegeeigenpertwo} only involves the two intermediate states $(n=3,m=2)$ and
$(n=2,m=3)$ with the unperturbed complex-conjugated eigenvalues
\begin{eqnarray}
 \lambda^{unper}_{n=3,m=2} && =-\frac{\Gamma+\Gamma'}{2}+i 2(h_2-h_1)
\nonumber \\
\lambda^{unper}_{n=2,m=3} && =-\frac{\Gamma+\Gamma'}{2}-i 2(h_2-h_1)
\label{unpertwo32}
\end{eqnarray}
and becomes
\begin{eqnarray}
\lambda_{0}(s) && = 
\frac{ < \lambda^{unper (L)}_{0,0}   \vert
{\cal L}^{per (1,2)} 
 \vert \lambda^{unper (R)}_{3,2}   >
< \lambda^{unper (L)}_{3,2}   \vert
{\cal L}^{per (1,2)} 
\vert \lambda^{unper (R)}_{0,0} > }
{  0  -  \lambda^{unper}_{3,2} } 
\nonumber \\
&& +
\frac{ < \lambda^{unper (L)}_{0,0}   \vert
{\cal L}^{per (1,2)} 
 \vert \lambda^{unper (R)}_{2,3}   >
< \lambda^{unper (L)}_{2,3}   \vert
{\cal L}^{per (1,2)} 
\vert \lambda^{unper (R)}_{0,0} > }
{  0 -  \lambda^{unper}_{2,3} } 
\nonumber \\
&& =  J_1^2  ( 1- e^{-s} ) \left[ e^s (1+\mu)(1-\mu') - (1-\mu)(1+\mu') \right] 
\frac{\Gamma+\Gamma'}{ \left(  \frac{\Gamma+\Gamma'}{2}\right)^2 + 4 (h_2-h_1)^2 }
\nonumber \\
&& =  \frac{D }{2}
\left[  (e^s-1) (1+\mu)(1-\mu') +(e^{-s}-1) (1-\mu)(1+\mu')  \right]
\label{nondegeeigenpertwoterms}
\end{eqnarray}
where we have introduced the notation
\begin{eqnarray}
D && \equiv
 2 J_1^2  
\frac{\Gamma+\Gamma'}{ \left(  \frac{\Gamma+\Gamma'}{2}\right)^2 + 4 (h_2-h_1)^2 }
\label{Dnota}
\end{eqnarray}

\subsection{ Averaged current and fluctuations }

The expansion of the eigenvalue of Eq. \ref{nondegeeigenpertwoterms}
up to second order in $s$ (Eq. \ref{eigensexpansion})
\begin{eqnarray}
\lambda_{0}(s) && = 
  \frac{D }{2}
\left[ s 2(\mu-\mu' ) +s^2 (1 -\mu \mu' )\right]
+O(s^3)
\label{nondegeeigenpertwotermsexp}
\end{eqnarray}
yields the averaged current (Eq \ref{siav})
\begin{eqnarray}
I_{av}  = \lim \limits_{t \to +\infty} \frac{ <N_t> }{ t} 
=  D (\mu-\mu' ) 
\label{siavtwo}
\end{eqnarray}
and the fluctuation (Eq \ref{sfluct})
\begin{eqnarray}
F  = \lim \limits_{t \to +\infty} \frac{ (<N_t^2>-<N_t>^2 ) }{ t} 
= D   (1 -\mu \mu' )
\label{sflucttwo}
\end{eqnarray}

\subsection{ Large deviations }

To compute the function $\Phi(I)$ that governs the large-deviation form of the probability distribution $P_t(I)$ of the current $I=\frac{N_t}{t}$ (Eq. \ref{larfedev}), we need the Legendre transform of Eq. \ref{legendre}
\begin{eqnarray}
  \Phi(I) = \opmax_s ( I s - \lambda_0(s) ) = I s_I - \lambda_0(s_I)
\label{legendrecol}
\end{eqnarray}
where $s_I$ is the location of the maximum determined by the solution of the equation
\begin{eqnarray}
  I = \lambda_0'(s)  && =
 \frac{D}{2} \left[  e^s (1+\mu)(1-\mu') - e^{-s} (1-\mu)(1+\mu')  \right]
\label{legendrederi}
\end{eqnarray}
Since this is a second-order equation in the variable $e^s$ 
\begin{eqnarray}
 0 =   (1+\mu)(1-\mu') e^{2s} - \frac{2 I }{D}  e^s-  (1-\mu)(1+\mu') 
\label{legendreeq}
\end{eqnarray}
with the discriminant 
\begin{eqnarray}
 \Delta = \left(\frac{2 I }{D}   \right)^2 +4 (1-\mu^2)(1-(\mu')^2)
\label{legendrediscri}
\end{eqnarray}
one obtains that the positive roots reads
\begin{eqnarray}
 e^{s_I} && = \frac{ \sqrt{  \left( \frac{I}{ D }\right)^2 +  (1-\mu^2) (1-(\mu')^2)  }   +\frac{I}{ D }   }
{   (1+\mu)(1-\mu')  } 
 = \frac{   (1-\mu)(1+\mu')  } { \sqrt{  \left( \frac{I}{ D }\right)^2 +  (1-\mu^2) (1-(\mu')^2)  }   -\frac{I}{ D }   }
\label{solusi}
\end{eqnarray}
so that the large deviation function of Eq. \ref{legendrecol} finally reads
\begin{eqnarray}
  \Phi(I)  && = I s_I -\frac{D }{2}
\left[  (e^{s_I}-1) (1+\mu)(1-\mu') +(e^{-s_I}-1) (1-\mu)(1+\mu')  \right]
\nonumber \\
&& = D (1-\mu \mu') 
-   \sqrt{  I^2 + D^2 (1-\mu^2) (1-(\mu')^2)  }  
+ I \ln \left[  \frac{ \sqrt{  I^2 + D^2 (1-\mu^2) (1-(\mu')^2)  }   +I  }
{  D (1+\mu)(1-\mu')  } \right]
\label{legendresol}
\end{eqnarray}
It vanishes $\Phi(I_{av}) =0 $ at $I_{av}=D(\mu-\mu')$ of Eq. \ref{siavtwo}
as it should.

\subsection{ Discussion }

In summary, besides the magnetizations $\mu$ and $\mu'$ of the boundary drivings,
the important parameter in the averaged current $I_{av}$,
 in the fluctuation $F$ and
more generally in the whole large-deviation function $\Phi(I)$
is the parameter $D$ introduced in Eq. \ref{Dnota} that contains the difference of the two random fields $(h_2-h_1) $
in the denominator.
In the remaining of the paper, we focus on the 'Strong-Disorder regime' where the scale of the random fields $h_j$
 is much bigger than the scale of the couplings $J_j$ that can be either uniform or random
\begin{eqnarray}
 (h_{j+1}-h_j)^2  \gg   J_j^2  
\label{strongh}
\end{eqnarray}
so that it is valid to use perturbation theory in the hoppings to evaluate various observables,
as shown in this section on the example of $N=2$ spins.

\section { Renormalization approach for the relaxation with a single reservoir}

\label{sec_onereservoir}

When the quantum chain of $N$ spins
 is subject to the single boundary-magnetization-driving (Eq \ref{dissiboundaryleftladder}) of 
parameters $(\Gamma,\mu)$ on the spin 1
(while there is no driving on the last spin $N$), the stationary state
is the trivial tensor-product with the magnetization $\mu$ for all spins
\begin{eqnarray}
\vert \lambda_{n=0}^R  > = \otimes_{j=1}^N \left( \frac{1+\mu}{2} \vert S_j=1,T_j=1> +\frac{1-\mu}{2} \vert S_j=-1,T_j=-1>  \right)
\label{trivialsteady}
\end{eqnarray}
but it is nevertheless interesting to analyze the behavior of 
the relaxation rate $\Gamma_N$ as a function of the system size $N$.

\subsection{ Boundary Strong Disorder Renormalization for the relaxation rate $\Gamma_N$ }

The idea is that in the Strong Disorder regime for the random fields (Eq. \ref{strongh}),
there exists a strong hierarchy between the relaxation rates
\begin{eqnarray}
  \Gamma_{N+1} \ll \Gamma_{N} \ll ... \ll \Gamma_1=\Gamma
\label{hierarchie}
\end{eqnarray}
i.e. the first spin $\sigma_1$ in contact with the reservoir is the first to equilibrate with rate $\Gamma_1=\Gamma$, then the second spin $\sigma_2$ will equilibrate with some slower rate $\Gamma_2$, and so on. 
The aim is thus to introduce a Boundary Strong Disorder Renormalization procedure
in order to compute iteratively the relaxation rates $\Gamma_{N}$.

So we decompose the Lindbladian for the chain of $(N+1)$ spins into
\begin{eqnarray}
{\cal L}_{N+1} && = {\cal L}^{unper}_{N+1}+{\cal L}^{per}_{N+1}
\nonumber \\
{\cal L}^{unper}_{N+1} && = {\cal L}_{N}+ i h_{N+1}(\tau_{N+1}^z - \sigma_{N+1}^z) 
\nonumber \\
{\cal L}^{per}_{N+1} && =
 i 2 J_N (\tau_N^+ \tau_{N+1}^-+\tau_N^- \tau_{N+1}^+ -\sigma_N^+ \sigma_{N+1}^--\sigma_N^- \sigma_{N+1}^+ ) 
\label{perturpationl}
\end{eqnarray}
in order to take into account the coupling term ${\cal L}^{per}_{N+1}  $ 
by perturbation theory in the hopping $J_N$.

\subsection{ Structure of the four lowest modes of ${\cal L}_{N} $   }

\label{4lowest}

When the strong hierarchy of Eq. \ref{hierarchie} exists,
one may restrict the Lindbladian ${\cal L}_{N} $ to its four lowest modes
\begin{eqnarray}
{\cal L}_N^{lowest} && = \sum_{n=0}^3  \lambda^{(N)}_i   \vert \psi^R_{\lambda^{(N)}_i} > <\psi^L_{\lambda^{(N)}_i} \vert   
\label{fourmodes}
\end{eqnarray}
that have the following structure for the last spin $N$ 
(while all the previous spins $j=1,..,N-1$ have already relaxed towards equilibrium) :

(0) The vanishing eigenvalue $\lambda^{(N)}_0=0$ representing the equilibrium
is associated to the left an right eigenvectors
\begin{eqnarray}
  < \lambda^{(N) L}_0 \vert  && =  < S_N=+,T_N=+ \vert + < S_N=-,T_N=- \vert
\nonumber \\
    \vert \lambda^{(N) R}_0  > && =  \frac{1+\mu}{2} \vert S_N=+,T_N=+> +\frac{1-\mu}{2} \vert S_N=-,T_N=- >  
\label{eigenrg0}
\end{eqnarray}

(1) The real eigenvalue $\lambda_1^{(N)}=-\Gamma_N$ is associated to 
\begin{eqnarray}
  < \lambda_1^{(N)L} \vert  && = \frac{1-\mu}{2} < S_N=+,T_N=+ \vert  - \frac{1+\mu}{2} <S_N=-,T_N=-\vert 
\nonumber \\
    \vert \lambda_1^{(N)R} > && = \vert S_N=+,T_N=+>  - \vert S_N=-,T_N=->
\label{eigenrg1}
\end{eqnarray}

(2-3) The complex eigenvalue $\lambda_2^{(N)}=-\frac{\Gamma_N}{2} +i (2 h_N+\omega_N)$ is associated to 
\begin{eqnarray}
  < \lambda_2^{(N)L} \vert  && = <  S_N=-,T_N=+\vert  
\nonumber \\
    \vert \lambda_2^{(N)R} > && = \vert S_N=-,T_N=+ >  
\label{eigenrg2}
\end{eqnarray}
while the complex-conjugate eigenvalue $\lambda_3^{(N)}=-\frac{\Gamma_N}{2}-i (2 h_N+\omega_N)$ is associated to 
\begin{eqnarray}
  < \lambda_3^{(N)L} \vert  && = <  S_N=+,T_N=- \vert  
\nonumber \\
    \vert \lambda_3^{(N)R} > && = \vert S_N=+,T_N=-  >  
\label{eigenrg3}
\end{eqnarray}

\subsection{ Properties of the unperturbed Lindbladian $ {\cal L}^{un}_{N+1} $ }

The lowest modes sector of the decoupled unperturbed Lindbladian reads
\begin{eqnarray}
{\cal L}^{unper}_{N+1} && = {\cal L}^{lowest}_{N}+ i h_{N+1}(\tau_{N+1}^z - \sigma_{N+1}^z) 
\nonumber \\
 && =  \sum_{n=0}^3  \lambda^{(N)}_n   \vert \lambda^{(N)R}_n > <\lambda^{(N)L}_n \vert  
 + i h_{N+1}(\tau_{N+1}^z - \sigma_{N+1}^z)
\label{unperturbed}
\end{eqnarray}
 so that its eigenstates are simply tensor-products
of eigenstates of each term
\begin{eqnarray}
{\cal L}^{unper}_{N+1}   \vert \lambda^{(N)R}_n > \otimes \vert S_{N+1},T_{N+1}> && =
\lambda^{(unper)}_{n,S_{N+1},T_{N+1}} \vert \lambda^{(N)R}_n > \otimes \vert S_{N+1},T_{N+1}>
\nonumber \\
< \lambda^{(N)L}_n \vert  \otimes < S_{N+1},T_{N+1} \vert {\cal L}^{unper}_{N+1} 
&& = \lambda^{(unper)}_{n,S_{N+1},T_{N+1}} < \lambda^{(N)L}_n \vert  \otimes < S_{N+1},T_{N+1} \vert
\label{eigenlzero}
\end{eqnarray}
and the corresponding eigenvalues are simply the sums
\begin{eqnarray}
\lambda^{(unper)}_{n,S_{N+1},T_{N+1}}= \lambda^{(N)}_n + i    h_{N+1}(T_{N+1} - S_{N+1}) 
\label{unpersumni}
\end{eqnarray}

In particular, the four eigenvalues corresponding to $n=0$ (with $S_{N+1}=\pm 1$ and $T_{N+1}=\pm 1$) have no real part as a consequence of $\lambda^{(N)}_{n=0}=0 $.
After taking into account the perturbation ${\cal L}^{per}_{N+1}  $ of Eq. \ref{perturpationl}, these four eigenvalues will correspond to the four slowest modes
of ${\cal L}_{N+1}$, with the structure analog to Eq. \ref{fourmodes}.

Since the perturbation has no diagonal contribution, we need to consider
the second-order perturbation theory for the eigenvalues.
Let us first consider the two complex-conjugate non-degenerate eigenvalues
\begin{eqnarray}
\lambda^{(unper)}_{0,+,-}= -  i 2 h_{N+1}
\nonumber \\
\lambda^{(unper)}_{0,-,+}= +  i 2 h_{N+1}
\label{nondege}
\end{eqnarray}
before we turn to the two-dimensional degenerate subspace
\begin{eqnarray}
\lambda^{(unper)}_{n=0,++}= \lambda^{(unper)}_{n=0,--} =0
\label{degesub}
\end{eqnarray}

\subsection{Second-Order Perturbation for
the two imaginary non-degenerate eigenvalues  }

In this section, we focus on the two imaginary complex-conjugate
 non-degenerate eigenvalues of Eq. \ref{nondege}.
The unperturbed eigenvalue 
\begin{eqnarray}
\lambda^{(unper)}_{0,+,-}= -  i 2 h_{N+1}
\label{nondegeeps}
\end{eqnarray}
corresponding to the left and right unperturbed eigenvectors (Eq. \ref{eigenlzero})
\begin{eqnarray}
< \lambda^{(unper)L}_{0,+,-} \vert 
&& 
= (< S_N=+,T_N=+  \vert + < S_N=-,T_N=-   \vert 
) \otimes <  S_{N+1}= + ,T_{N+1}=- \vert 
\\
\vert \lambda^{(unper)R}_{0,+,-}> && 
= \left( \frac{1+\mu}{2} \vert S_N=+,T_N=+ > 
+\frac{1-\mu}{2} \vert S_N=-,T_N=-  >  \right)
\otimes \vert  S_{N+1}= + ,T_{N+1}=- >
\nonumber 
\label{leftrightzero}
\end{eqnarray}
has the following second-order perturbation correction 
\begin{eqnarray}
\lambda^{(2^dorder)}_{0,+,-}= 
\sum_{(n,S,T) \ne (0,+,- )} 
\frac{ <  \lambda^{(unper)L}_{0,+,-} \vert {\cal L}^{(per)}_{N+1} 
 \vert \lambda^{(unper)R}_{n,S,T} >
<  \lambda^{(unper)L}_{n,S,T} \vert
 {\cal L}^{(per)}_{N+1}  \vert \lambda^{(unper)R}_{0,+,-}> }{\lambda^{(unper)}_{0,+,-}-\lambda^{(unper)}_{n,S,T} } 
\label{nondegeeigenper}
\end{eqnarray}

The application of the perturbation ${\cal L}^{(per)}_{N+1} $ of Eq. \ref{perturpationl}
on the left and right eigenvectors yield 
\begin{eqnarray}
< \lambda^{(unper)L}_{0,+,-} \vert {\cal L}^{(per)}_{N+1}  
&& 
= i 2 J_N
 (< \lambda^{(unper)L}_{3,+,+}  \vert - < \lambda^{(unper)L}_{3,-,-}  \vert  ) 
\nonumber  \\
{\cal L}^{(per)}_{N+1}\vert \lambda^{(unper)R}_{0,+,-}> 
&& 
= i 2 J_N  \left( \frac{1+\mu}{2} \vert \lambda^{(unper)R}_{3,+,+} >
-\frac{1-\mu}{2} \vert \lambda^{(unper)R}_{3,-,-} >\right)
\label{allpliperleftright}
\end{eqnarray}

So the sum of Eq. \ref{nondegeeigenper} contains only two terms corresponding to the unperturbed eigenvalues
\begin{eqnarray}
\lambda^{(unper)}_{3,+,+}= \lambda^{(unper)}_{3,-,-} = \lambda^{(N)}_3 =-\frac{\Gamma_N}{2} -i (2 h_N+\omega_N)
\label{eigenn3}
\end{eqnarray}
and finally reads
\begin{eqnarray}
\lambda^{(2^dorder)}_{0,+,-} && = 
\frac{ <  \lambda^{(unper)L}_{0,+,-} \vert {\cal L}^{(per)}_{N+1}  \vert \lambda^{(unper)R}_{3,+,+} >
<  \lambda^{(unper)L}_{3,+,+} \vert {\cal L}^{(per)}_{N+1}  \vert \lambda^{(unper)R}_{0,+,-}> }
{\lambda^{(unper)}_{0,+,-}-\lambda^{(unper)}_{3,+,+} } 
\nonumber \\
&& + \frac{ <  \lambda^{(unper)L}_{0,+,-} \vert {\cal L}^{(per)}_{N+1}  \vert \lambda^{(unper)R}_{3,-,-} >
<  \lambda^{(unper)L}_{3,-,-} \vert {\cal L}^{(per)}_{N+1}  \vert \lambda^{(unper)R}_{0,+,-}> }
{\lambda^{(unper)}_{0,+,-}-\lambda^{(unper)}_{3,-,-} } 
\nonumber \\
&& = - \frac{ 4 J_N^2 }{\frac{\Gamma_N}{2} +i (2 h_N+\omega_N -2 h_{N+1}) }
\label{nondegeeigenperres}
\end{eqnarray}

For the other complex-conjugate unperturbed eigenvalue
\begin{eqnarray}
\lambda^{(unper)}_{0,-,+}= +  i 2 h_{N+1}
\label{nondegeepsbis}
\end{eqnarray}
the second-order perturbation is similar and yields of course
the complex-conjugate result of Eq. \ref{nondegeeigenperres}
\begin{eqnarray}
\lambda^{(2^dorder)}_{0,-,+} 
&& =  - \frac{ 4 J_N^2 }{\frac{\Gamma_N}{2} -i (2 h_N+\omega_N -2 h_{N+1}) }
\label{nondegeeigenperrescc}
\end{eqnarray}

In summary, the identification of these two complex-conjugate eigenvalues
\begin{eqnarray}
-\frac{\Gamma_{N+1}}{2} +i (2 h_{N+1}+\omega_{N+1}) = \lambda^{unper}_{0,-,+}
+ \lambda^{2^d order}_{0,-,+} && = i 2 h_{N+1} - \frac{ 4 J_N^2 }{\frac{\Gamma_N}{2} +i (2 h_N+\omega_N -2 h_{N+1}) }
\nonumber \\
-\frac{\Gamma_{N+1}}{2} -i (2 h_{N+1}+\omega_{N+1}) = \lambda^{unper}_{0,+,-} + \lambda^{2^d order}_{0,+,-}&& = -i 2 h_{N+1} - \frac{ 4 J_N^2 }{\frac{\Gamma_N}{2} +i (2 h_N+\omega_N -2 h_{N+1}) }
\label{identificationrg}
\end{eqnarray}
leads to the following recurrences for the two variables $(\Gamma_N,\omega_N)$
\begin{eqnarray}
\Gamma_{N+1}  = \frac{ 4 J_N^2 \Gamma_N }{ (\frac{\Gamma_N}{2})^2 + (2 h_N+\omega_N -2 h_{N+1})^2 }
\label{recgreal}
\end{eqnarray}
and
\begin{eqnarray}
 \omega_{N+1} = \frac{ 4 J_N^2 (2 h_N+\omega_N -2 h_{N+1})   }{ (\frac{\Gamma_N}{2})^2 + (2 h_N+\omega_N -2 h_{N+1})^2 }
\label{recgimag}
\end{eqnarray}

\subsection{ Perturbation in the two-dimensional degenerate subspace
 $\lambda^{(unper)}_{n=0,++}= \lambda^{(unper)}_{n=0,--} =0$  }

Within the two-dimensional degenerate subspace
 $\lambda^{(unper)}_{n=0,++}= \lambda^{(unper)}_{n=0,--} =0$ associated to the projector
\begin{eqnarray}
{\cal P}_0   = 
\vert \lambda^{(unper)R}_{n=0,++} >  <  \lambda^{(unper)L}_{n=0,++}\vert 
+ \vert \lambda^{(unper)R}_{n=0,--} >  <  \lambda^{(unper)L}_{n=0,--}\vert 
\label{proj0}
\end{eqnarray}
the second-order perturbation theory corresponds to the effective operator
(that generalizes the non-degenerate perturbation formula of Eq. \ref{nondegeeigenper})
\begin{eqnarray}
{\cal L}^{(2^dorder)} _{\lambda=0}=  {\cal P}_0 {\cal L}^{(per)}_{N+1} (1-{\cal P}_0 )\frac{1}{0- {\cal L}^{(unper)}_{N+1}} 
(1-{\cal P}_0 ){\cal L}^{(per)}_{N+1} {\cal P}_0
\label{Leff2emeordre}
\end{eqnarray}

The application of the perturbation ${\cal L}^{(per)}_{N+1} $ of Eq. \ref{perturpationl}
on the left and right eigenvectors yield respectively
\begin{eqnarray}
< \lambda^{(unper)L}_{0,+,+} \vert {\cal L}^{(per)}_{N+1}
&& = i 2 J_N
 (< \lambda^{(unper)L}_{2,+,-}  \vert - < \lambda^{(unper)L}_{3,-,+}  \vert  ) 
\nonumber  \\
{\cal L}^{(per)}_{N+1}\vert \lambda^{(unper)R}_{0,+,+}> 
&& 
= i 2 J_N  \frac{1-\mu}{2} \left( \vert \lambda^{(unper)R}_{2,+,-} >
- \vert \lambda^{(unper)R}_{3,-,+} >\right)
\label{appliperleft}
\end{eqnarray}

and

\begin{eqnarray}
< \lambda^{(unper)L}_{0,-,-} \vert {\cal L}^{(per)}_{N+1}
&& 
= i 2 J_N
 (-< \lambda^{(unper)L}_{2,+,-}  \vert + < \lambda^{(unper)L}_{3,-,+}  \vert  ) 
\nonumber  \\
{\cal L}^{(per)}_{N+1}\vert \lambda^{(unper)R}_{0,-,-}> 
&& 
= i 2 J_N  \frac{1+\mu}{2} \left( - \vert \lambda^{(unper)R}_{2,+,-} >
+ \vert \lambda^{(unper)R}_{3,-,+} >\right)
\label{appliperright}
\end{eqnarray}

As a consequence, Eq. \ref{Leff2emeordre} contains
 only two intermediate states 
that are associated to the unperturbed eigenvalues
\begin{eqnarray}
\lambda^{(unper)}_{2,+-}= \lambda^{(N)}_2 - i 2   h_{N+1} = - \frac{\Gamma_N}{2} 
+i (2 h_N+\omega_N-2   h_{N+1} )
\nonumber \\
\lambda^{(unper)}_{3,-+}= \lambda^{(N)}_3 + i 2   h_{N+1} = - \frac{\Gamma_N}{2}
 -i (2 h_N+\omega_N-2   h_{N+1} )
\label{unperinterdege}
\end{eqnarray}
and becomes
\begin{eqnarray}
{\cal L}^{(2^dorder)} _{\lambda=0}=  {\cal P}_0 {\cal L}^{(per)}_{N+1} 
\left[ \frac{ \vert\lambda^{(unper)R}_{2,+-} > <  \lambda^{(unper)L}_{2,+-}\vert}{0- \lambda^{(unper)}_{2,+-}} 
+\frac{ \vert\lambda^{(unper)R}_{3,-+} > <  \lambda^{(unper)L}_{3,-+}\vert}{0- \lambda^{(unper)}_{3,-+}} 
\right]  {\cal L}^{(per)}_{N+1} {\cal P}_0
\label{Leff2emeordreinter}
\end{eqnarray}

In terms of the notation $\Gamma_{N+1}$ introduced in Eq. \ref{recgreal},
the four corresponding matrix elements read

\begin{eqnarray}
  <  \lambda^{(unper)L}_{n=0,++}\vert {\cal L}^{(2^dorder)} _{\lambda=0} \vert \lambda^{(unper)R}_{n=0,++} > 
&& = -  \frac{1-\mu}{2} \Gamma_{N+1}
\label{matrixel1}
\end{eqnarray}

\begin{eqnarray}
  <  \lambda^{(unper)L}_{n=0,--}\vert {\cal L}^{(2^dorder)} _{\lambda=0} \vert \lambda^{(unper)R}_{n=0,--} > 
&& = -  \frac{1+\mu}{2} \Gamma_{N+1}
\label{matrixel2}
\end{eqnarray}

\begin{eqnarray}
  <  \lambda^{(unper)L}_{n=0,++}\vert {\cal L}^{(2^dorder)} _{\lambda=0} \vert \lambda^{(unper)R}_{n=0,--} > 
&& =  \frac{1+\mu}{2} \Gamma_{N+1}
\label{matrixel3}
\end{eqnarray}

\begin{eqnarray}
  <  \lambda^{(unper)L}_{n=0,--}\vert {\cal L}^{(2^dorder)} _{\lambda=0} \vert \lambda^{(unper)R}_{n=0,++} > 
&& =  \frac{1-\mu}{2} \Gamma_{N+1}
\label{matrixel4}
\end{eqnarray}

So the two-by-two matrix can be factorized into
\begin{eqnarray}
{\cal L}^{(2^dorder)} _{\lambda=0}= - \Gamma_{N+1} 
\left( \vert \lambda^{(unper)R}_{n=0,++} >- \vert \lambda^{(unper)R}_{n=0,--} > \right)
\left( \frac{1-\mu}{2}  <  \lambda^{(unper)L}_{n=0,++}\vert  -  \frac{1+\mu}{2} <  \lambda^{(unper)L}_{n=0,--}\vert\right)
\label{Leff2emeordrefactor}
\end{eqnarray}
where
\begin{eqnarray}
\lambda^{(N+1)}_{n=1}=- \Gamma_{N+1} = - 4 J_N^2  \frac{\Gamma_N}{(\frac{\Gamma_N}{2})^2 + ((2 h_N+\omega_N) -2 h_{N+1})^2}
\label{Leff2emeordrepp}
\end{eqnarray}
represents the eigenvalue associated to the right and left eigenvectors
\begin{eqnarray}
\vert \lambda^{(N+1)R}_{n=1} >
&& =
 \vert \lambda^{(unper)R}_{n=0,++} >- \vert \lambda^{(unper)R}_{n=0,--} > 
\nonumber \\
<\lambda^{(N+1)L}_{n=1} \vert =
&&
 \frac{1-\mu}{2}  <  \lambda^{(unper)L}_{n=0,++}\vert  -  \frac{1+\mu}{2} <  \lambda^{(unper)L}_{n=0,--}\vert
\label{eigennext1}
\end{eqnarray}
while the vanishing eigenvalue $\lambda^{N+1}_{n=0}=0$ corresponds to the right and left eigenvectors
\begin{eqnarray}
\vert \lambda^{(N+1)R}_{n=0} >
&& =
  \frac{1+\mu}{2} \vert \lambda^{(unper)R}_{n=0,++} >
 + \frac{1-\mu}{2}  \vert  \lambda^{(unper)R}_{n=0,--}>
\nonumber \\
<\lambda^{(N+1)L}_{n=0} \vert =
&&
 < \lambda^{(unper)L}_{n=0,++}\vert 
+ < \lambda^{(unper)L}_{n=0,--}\vert 
\label{eigennext0}
\end{eqnarray}

\subsection{ Validity of the Strong Disorder Approach  }

In summary, we have obtained the recurrences of Eqs \ref{recgreal}, \ref{recgimag} 
for the two variables $(\Gamma_N,\omega_N)$ that 
characterize the structure of the four slowest modes described in sec \ref{4lowest}.
The above perturbative calculation is valid in the strong disorder regime for the random fields (Eq. \ref{strongh}) that leads to a strong hierarchy between the relaxation rates
(Eq. \ref{hierarchie}). When this is the case, the relaxation rate $\Gamma_N$ decays with $N$
and can be neglected with respect to the difference of random fields $(h_N-h_{N+1})$
in the denominators of the recurrences of Eqs \ref{recgreal} and \ref{recgimag} that becomes
\begin{eqnarray}
\Gamma_{N+1}  \simeq \frac{ 4 J_N^2  }{ (2 h_N+\omega_N -2 h_{N+1})^2 }\Gamma_N
\label{recgrealstrong}
\end{eqnarray}
and
\begin{eqnarray}
 \omega_{N+1} \simeq \frac{ 4 J_N^2    }{  2 h_N+\omega_N -2 h_{N+1} }
\label{recgimagstrong}
\end{eqnarray}
At leading order in the strong disorder regime for the random fields,
one further obtains that $\omega_{N}$ can be neglected
 with respect to the difference of random fields $(h_N-h_{N+1})$
in the denominators leading to the simple value for the correction to the imaginary part
\begin{eqnarray}
 \omega_{N+1} \simeq \frac{ 2 J_N^2    }{   h_N - h_{N+1} }
\label{recgimagstrongleading}
\end{eqnarray}
and to the simplified multiplication recurrence for the relaxation rates alone
\begin{eqnarray}
\Gamma_{N+1}  \simeq \frac{  J_N^2  }{ ( h_N - h_{N+1})^2 }\Gamma_N
\label{recgrealstrongleading}
\end{eqnarray}
This result clearly shows that the hypothesis of Eq. \ref{hierarchie} 
concerning the strong hierarchy between two sucessive relaxation rates $\Gamma_{N+1} \ll \Gamma_N$
is satisfied in the strong disorder regime  $ ( h_N - h_{N+1})^2  \gg J_N^2$ (Eq. \ref{strongh}),
so that the renormalization procedure described in the present section is fully consistent.

In terms of the initial relaxation rate $\Gamma_1=\Gamma$ and of the random fields $h_j$ and random couplings $J_j$, the relaxation rate $\Gamma_N$ 
is then simply given by the product
\begin{eqnarray}
\Gamma_{N}  \simeq \Gamma \prod_{j=1}^{N-1} \frac{  J_j^2  }{ ( h_j - h_{j+1})^2 }
\label{gammanprod}
\end{eqnarray}
so that its logarithm corresponds to a sum of independent random variables
\begin{eqnarray}
\ln \Gamma_{N}  \simeq \ln \Gamma+\sum_{j=1}^{N-1} \ln \left( \frac{  J_j^2  }{ ( h_j - h_{j+1})^2 } \right)
\label{gammanprodln}
\end{eqnarray}
The Central Limit Theorem then yields that the distribution of $(\ln \Gamma_{N} )$ over the disordered samples
  is Gaussian
with the average
\begin{eqnarray}
\overline{ \ln \Gamma_{N} }  \simeq \overline{\ln \Gamma_1}
+(N-1) \ \overline{ \ln \left( \frac{  J_j^2  }{ ( h_j - h_{j+1})^2 } \right) }
\label{gammanprodlnav}
\end{eqnarray}
and the variance
\begin{eqnarray}
{\rm Var} [ \ln \Gamma_{N} ]  \simeq (N-1) {\rm Var}  \left[ \ln \left( \frac{  J_j^2  }{ ( h_j - h_{j+1})^2 } \right)  \right]
\label{gammanprodlnvar}
\end{eqnarray}

\section{ Non-equilibrium steady states between two reservoirs }

\label{sec_tworeservoirs}

\subsection{ Non-equilibrium magnetization profile between two reservoirs }

Although the simplest expectation for a non-equilibrium steady-state between two reservoirs 
would be a linear magnetization profile as in the Fourier-Fick-diffusive standard result,
it should be stressed that the completely opposite situation
 of a step-magnetization profile with a 'shock'
has been found in various regimes \cite{casati_step,clark_step,prosen_step}
and in particular in the presence of disorder as a consequence of 
the localization phenomenon \cite{huveneers_mblstep}.
In the strong disorder regime that we consider,
we also expect that the magnetization profile will have a step-profile :
the magnetization will remain near $\mu$ for the spins $j=1,2,..,n$,
while it will remain near $\mu'$ for the other spins $j=n+1,..,N$.
In this section, our goal is to determine the location $(n,n+1)$ of the step
as a function of the random fields of the sample.
This step magnetization profile means that the reservoir acting on the spin 1 is actually able
to impose its magnetization $\mu$ on all the spins $j=1,2,..,n$,
while the other reservoir acting on the spin N is actually able
to impose its magnetization $\mu'$ on all the spins $j=n+1,..,N$,
so that we may directly use the results of the previous section 
concerning the relaxation in the presence of a single reservoir :

(i) For the spin $n$, the effective Lindbladian describing the influence of the left reservoir acting on spin 1
reads in terms of the four states $\vert S_n=\pm 1, T_n=\pm 1>$ of the ladder formulation

\begin{eqnarray}
{\cal L}^{Left} _{ n} &&
= 0 \times 
\left(\frac{1+\mu}{2} \vert ++ >+ \frac{1-\mu}{2} \vert  -- > \right)
\left(   < ++ \vert  -+ <  -- \vert\right)
\nonumber \\
&& - \Gamma^{Left}_{n} 
\left( \vert ++ >- \vert  -- > \right)
\left( \frac{1-\mu}{2}  < ++ \vert  -  \frac{1+\mu}{2} <  -- \vert\right)
\nonumber \\
&& - \frac{\Gamma^{Left}_{n} }{2} \vert +- > < +- \vert  
\nonumber \\
&& -  \frac{\Gamma^{Left}_{n} }{2} \vert -+> < -+ \vert  
\label{leftleff}
\end{eqnarray}
with the relaxation rate given by Eq. \ref{gammanprod}
\begin{eqnarray}
\Gamma^{Left}_{n}  \simeq \Gamma \prod_{j=1}^{n-1} \frac{  J_j^2  }{ ( h_j - h_{j+1})^2 }
\label{gammanprodleft}
\end{eqnarray}

(ii) Similarly, for the spin $n$, the effective Linbladian describing the influence of the right reservoir acting on spin $N$ reads

\begin{eqnarray}
&& {\cal L}^{Right} _{ n} 
= 0 
\left(\frac{1+\mu'}{2} \vert ++ >+ \frac{1-\mu'}{2} \vert  -- > \right)
\left(   < ++ \vert  -+ <  -- \vert\right)
\nonumber \\
&& - \Gamma^{Right}_{n} 
\left( \vert ++ >- \vert  -- > \right)
\left( \frac{1-\mu'}{2}  < ++ \vert  -  \frac{1+\mu'}{2} <  -- \vert\right)
\nonumber \\
&& - \frac{\Gamma^{Right}_{n} }{2} \vert +- > < +- \vert  
\nonumber \\
&& -  \frac{\Gamma^{Right}_{n} }{2} \vert -+> < -+ \vert  
\label{rightleff}
\end{eqnarray}
with the relaxation rate given by the appropriate adaptation of Eq.
\ref{gammanprod}
\begin{eqnarray}
\Gamma^{Right}_{n}  \simeq \Gamma' \prod_{j=n}^{N-1} \frac{  J_j^2  }{ ( h_j - h_{j+1})^2 }
\label{gammanprodright}
\end{eqnarray}

(iii) Taking into account the random field $h_n$, one finally obtain that the total
effective Lindbladian acting on the spin $n$ reads
\begin{eqnarray}
{\cal L}^{tot} _{ n} && = {\cal L}^{Left} _{ n}+ {\cal L}^{Right}_{n} + i 2 h_n (\vert +- > < +- \vert  - \vert -+> < -+ \vert  ) 
\nonumber \\
&& = -
\left( \vert ++ >- \vert  -- > \right)
\nonumber \\
&& 
\left( \left[\Gamma^{Left}_{n} \frac{1-\mu}{2} +\Gamma^{Right}_{n}  \frac{1-\mu'}{2} \right] < ++ \vert
  - \left[ \Gamma^{Left}_{n} \frac{1+\mu}{2}+ \Gamma^{Right}_{n}  \frac{1+\mu'}{2} \right] <  -- \vert\right)
\nonumber \\
&& - \left( \frac{\Gamma^{Left}_{n}+\Gamma^{Right}_{n}  }{2}+ i 2 h_n \right) \vert +- > < +- \vert  
\nonumber \\
&& -  \left( \frac{\Gamma^{Left}_{n}+ \Gamma^{Right}_{n} }{2}- i 2 h_n \right) \vert -+> < -+ \vert  
\label{totleff}
\end{eqnarray}

So the global relaxation rate corresponds to the sum
\begin{eqnarray}
\Gamma_n^{tot} && =\Gamma^{Left}_{n}+\Gamma^{Right}_{n} 
\nonumber \\
&& = \Gamma \prod_{j=1}^{n-1} \frac{  J_j^2  }{ ( h_j - h_{j+1})^2 }
+ \Gamma' \prod_{j=n}^{N-1} \frac{  J_j^2  }{ ( h_j - h_{j+1})^2 }
\label{totgamman}
\end{eqnarray}
while the effective magnetization $\mu_n$ that the combination of the two reservoirs tend to impose on site $n$
can be obtained from the identification 
\begin{eqnarray}
\Gamma_n^{tot} \frac{1-\mu_n}{2} = \Gamma^{Left}_{n} \frac{1-\mu}{2} +\Gamma^{Right}_{n}  \frac{1-\mu'}{2} 
\nonumber \\
\Gamma_n^{tot} \frac{1+\mu_n}{2} = \Gamma^{Left}_{n} \frac{1+\mu}{2}+ \Gamma^{Right}_{n}  \frac{1+\mu'}{2} 
\label{totmunidentif}
\end{eqnarray}
 leading to the weighted average
\begin{eqnarray}
 \mu_n && =\frac{ \mu  \Gamma^{Left}_{n}  + \mu' \Gamma^{Right}_{n}  }{ \Gamma^{Left}_{n}+\Gamma^{Right}_{n}}
 =\frac{\displaystyle \mu \Gamma \prod_{j=1}^{n-1} \frac{  J_j^2  }{ ( h_j - h_{j+1})^2 } 
  + \mu' \Gamma' \prod_{j=n}^{N-1} \frac{  J_j^2  }{ ( h_j - h_{j+1})^2 }  }
{\displaystyle \Gamma \prod_{j=1}^{n-1} \frac{  J_j^2  }{ ( h_j - h_{j+1})^2 }
+ \Gamma' \prod_{j=n}^{N-1} \frac{  J_j^2  }{ ( h_j - h_{j+1})^2 } }
\label{totmun}
\end{eqnarray}

The same approach for the other spin $(n+1)$ on the other side of the step yields
\begin{eqnarray}
 \mu_{n+1} 
&& =\frac{\displaystyle  \mu \Gamma \prod_{j=1}^{n} \frac{  J_j^2  }{ ( h_j - h_{j+1})^2 } 
  + \mu' \Gamma' \prod_{j=n+1}^{N-1} \frac{  J_j^2  }{ ( h_j - h_{j+1})^2 }  }
{\displaystyle  \Gamma \prod_{j=1}^{n} \frac{  J_j^2  }{ ( h_j - h_{j+1})^2 }
+ \Gamma' \prod_{j=n+1}^{N-1} \frac{  J_j^2  }{ ( h_j - h_{j+1})^2 } }
\label{totmunnext}
\end{eqnarray}
This step magnetization profile approximation will be valid if
\begin{eqnarray}
 \mu_n \simeq \mu
\nonumber \\
 \mu_{n+1}\simeq \mu'
\label{steplocation}
\end{eqnarray}
and the location $(n,n+1)$ of the step correspond to the location where there is a change of the dominant reservoir in the weighted average.
For instance, for the standard example of opposite boundary magnetizations
 $\mu=-\mu' >0$ and equal boundary-rates $\Gamma'=\Gamma$, the location of the step corresponds to the index $n$ where
there is a sign change in the difference 
\begin{eqnarray}
 \prod_{j=1}^{n-1} \frac{  J_j^2  }{ ( h_j - h_{j+1})^2 }
-  \prod_{j=n}^{N-1} \frac{  J_j^2  }{ ( h_j - h_{j+1})^2 }
>0 > \prod_{j=1}^{n} \frac{  J_j^2  }{ ( h_j - h_{j+1})^2 }
-  \prod_{j=n+1}^{N-1} \frac{  J_j^2  }{ ( h_j - h_{j+1})^2 }
\label{steplocationzero}
\end{eqnarray}
So while the average position of the step is at the middle of the chain by symmetry,
there are sample-to-sample fluctuations of order $\sqrt{N}$ as a consequence of 
the statistical properties discussed after Eq. \ref{gammanprod}.

\subsection{ Non-equilibrium magnetization current between two reservoirs }

Within the picture of the step-magnetization-profile located on the bond $(n,n+1)$ described above,
the analysis of the current is actually similar to the two-spin problem described in detail in section \ref{sec_twospins}.
The important parameter of Eq. \ref{Dnota} becomes
\begin{eqnarray}
D_n && =
 2 J_n^2  
\frac{\Gamma_n^{Left}+\Gamma_{n+1}^{Right} }
{ \left(  \frac{\Gamma_n^{Left}+\Gamma_{n+1}^{Right}}{2}\right)^2 + 4 (h_{n+1}-h_n)^2 }
\label{Dnotan}
\end{eqnarray}
in terms of the relaxation rates $\Gamma_n^{Left} $ and $\Gamma_{n+1}^{Right} $ 
given by Eqs. \ref{gammanprodleft} and \ref{gammanprodright}.
Since they are small, they can be neglected in the denominator with respect to the random fields,
so that the parameter $D_n$ reads at leading order in the strong disorder regime
\begin{eqnarray}
D_n && =   \frac{  J_n^2 }
{ 2 (h_{n+1}-h_n)^2 }  \left( \Gamma_n^{Left}+\Gamma_{n+1}^{Right} \right)
\nonumber \\
&& =    \frac{  J_n^2 }
{ 2 (h_{n+1}-h_n)^2 }  \left( \Gamma \prod_{j=1}^{n-1} \frac{  J_j^2  }{ ( h_j - h_{j+1})^2 }
+ \Gamma' \prod_{j=n+1}^{N-1} \frac{  J_j^2  }{ ( h_j - h_{j+1})^2 }\right)
\label{Dnotafinal}
\end{eqnarray}
where the location $(n,n+1)$ of the step has been discussed after Eq. \ref{steplocation} : the two products in the parenthesis are then roughly of the same order.
As a consequence, the averaged current $I_{av}$ and the fluctuation $F$ given by Eqs \ref{siavtwo} and \ref{sflucttwo}
in terms of this parameter $D_n$
\begin{eqnarray}
I_{av}  && = \lim \limits_{t \to +\infty} \frac{ <N_t> }{ t} 
=  D_n (\mu-\mu' ) 
\nonumber \\
F  && = \lim \limits_{t \to +\infty} \frac{ (<N_t^2>-<N_t>^2 ) }{ t} 
= D_n   (1 -\mu \mu' )
\label{sflucttwon}
\end{eqnarray}
will be typically exponentially small with respect to the system-size $N$.
The probability distribution of $D_n$ over the samples is expected to be log-normal
 as a consequence of the product-structure discussed after Eq. \ref{gammanprod}.

\section{ Conclusions }

\label{sec_conclusion}

In this paper, we have considered the Lindblad dynamics of the XX quantum chain with large random fields $h_j$, while the couplings $J_j$ can be either uniform or random, for boundary-magnetization-drivings acting on the two end-spins. We have first analyzed the relaxation properties in the presence of a single reservoir as a function of the system size via some boundary-strong-disorder renormalization approach.
We have then studied the non-equilibrium-steady-state in the presence of two reservoirs via the effective renormalized Linbladians associated to the two reservoirs. The magnetization has been found to follow a step profile, as found previously in other localized chains \cite{huveneers_mblstep}. The strong disorder approach has been used to compute explicitly the location of the step of the magnetization profile and the corresponding exponentially-small magnetization-current for each disordered sample in terms of the random fields and couplings.

The companion paper \cite{c_dephasing} describes how the addition of bulk-dephasing in the dissipative part of the Linbladian destroys these localization properties.

\end{document}